\documentclass[twocolumn]{el-author}
\usepackage{multirow}
\usepackage[misc]{ifsym}

\newcommand{\ua}{\uparrow}
\newcommand{\nc}{\newcommand}
\nc{\da}{\downarrow} \nc{\hc}{\hat{c}} \nc{\hS}{\hat{S}}
\nc{\bra}{\langle} \nc{\ket}{\rangle} \nc{\eq}{equation (\ref}
\nc{\h}{\hat} \nc{\hT}{\h{T}}\nc{\be}{\begin{eqnarray}}
\nc{\ee}{\end{eqnarray}}\nc{\rd}{\textrm{d}}\nc{\e}{eqnarray}\nc{\hR}{\hat{R}}\nc{\Tr}{\mathrm{Tr}}
\nc{\tS}{\tilde{S}}\nc{\tr}{\mathrm{tr}}\nc{\8}{\infty}\nc{\lgs}{\bra\ua,\phi|}\nc{\rgs}{|\ua,\phi\ket}
\nc{\hU}{\hat{U}}\nc{\lfs}{\bra\phi|}\nc{\rfs}{|\phi\ket}\nc{\hZ}{\hat{Z}}\nc{\hd}{\hat{d}}\nc{\mD}{\mathcal{D}}
\nc{\bd}{\bar{d}}\nc{\bc}{\bar{c}}\nc{\mc}{\mathcal}\nc{\ea}{eqnarray}\nc{\mG}{\mathcal{G}}\nc{\bce}{\begin{center}}
\nc{\ece}{\end{center}}
\nc{\etal}{\textit{et al.}}

\begin{document}

\title{Power pooling: An adaptive pooling function for weakly labelled sound event detection}

\author{Yuzhuo Liu, Hangting Chen, Yun Wang and Pengyuan Zhang\textsuperscript{\Letter}}

\abstract{Access to large corpora with strongly labelled sound events is expensive and difficult in engineering applications. Much research turns to address the problem of how to detect both the types and the timestamps of sound events with weak labels that only specify the types. This task can be treated as a multiple instance learning (MIL) problem, and the key to it is the design of a pooling function. The state-of-the-art linear softmax pooling function, however, cannot flexibly deal with sound sources of different time scales. In this paper, we propose an adaptive power pooling function which can automatically adapt to various sound sources. On two public datasets, the proposed power pooling function outperforms linear softmax pooling on both coarse-grained and fine-grained metrics. Notably, it improves the event-based $F_1$ score by 11.4\% and 10.2\% relative on the two datasets. While this paper focuses on sound event detection applications, the proposed method can be applied to MIL tasks in other domains.}

\maketitle

\section{Introduction}

Sound event detection (SED) aims to identify the categories and timestamps of target sound events in continuous audio recordings. Some studies only focus on the categories of sound events present (\emph{audio tagging}), while this paper pays more attention to the detection of onsets and offsets of sound events (\emph{localization}). Traditional SED models are often trained from data with strong labels, which contain the categories and timestamps of each sound event occurrence~\cite{DNN, CNN, RNN, CRNN}. However, in real-world applications, such as noise monitoring, surveillance systems, machine condition monitoring, and multimedia indexing, acquiring such strong labels can incur a high cost. In 2017, Google released a large-scale weakly labelled dataset (AudioSet)~\cite{audioset} with annotations of only event categories at a coarse time resolution (10~seconds). AudioSet has led researches to pay more attention to weakly labelled SED, \textit{i.e.} when no fine-grained timestamps are available.

Weakly labelled SED can be addressed as a multiple instance learning (MIL) problem. An audio clip and the frames in it can be regarded as a bag and instances in the bag. A positive bag is a clip that contains a certain event; it consists of at least one positive frame and may also contain negative frames. A negative bag, on the other hand, consists only of negative frames. For improving the localization accuracy, considerable researches have made efforts to select positive instances more precisely. For weakly supervised object detection, Wan \etal~\cite{C-MIL} introduced a continuation optimization algorithm. Yang \etal~\cite{E-E} proposed to jointly train a MIL branch and a bounding-box regression branch. These methods activate more positive samples by avoiding falling into local minima. Cheng \etal~\cite{H-Q} focused on generating and selecting high-quality proposals to better envelop all positive samples.

\begin{figure}[t]
\centering{\includegraphics[width=80mm]{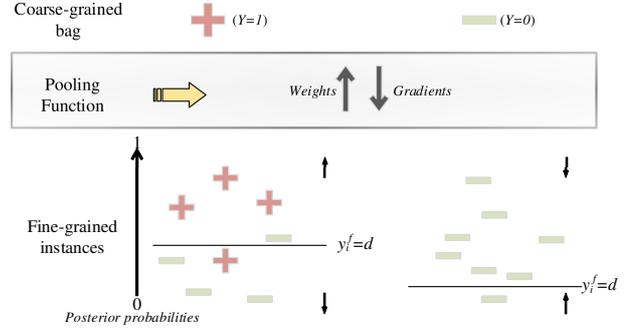}}
\caption{Pooling function in a MIL system for SED with weak labels. A pooling function produces weights for fine-grained prediction to obtain a coarse-grained prediction, and generates fine-grained gradients from a coarse-grained loss. The red plus signs indicate samples whose ground truth is positive, and the green minus signs indicate samples whose ground truth is negative. $d$ is the ideal threshold at which the gradients change sign. Arrows indicate the directions of the gradients.}
\label{fig:intro}
\source{}
\end{figure}

As for SED, many studies are devoted to the design of pooling functions. A MIL system predicts a probability for each frame, and aggregates the frame-level predictions into a clip-level probability via a \emph{pooling function}. As shown in Fig.~\ref{fig:intro}, the pooling function calculates the clip-level probability as a weighted average of the frame-level probabilities, and also serves to back-propagate gradients from the clip-level loss function to the frames. Ideally, the pooling function should be discriminative enough to produce positive gradients for positive frames and negative gradients for negative frames. \emph{``Max'' pooling} assigns zero weights to non-maximizing frames; this produces zero gradients and leads to difficult optimization. \emph{Average pooling}~\cite{mean} weights all frames equally and produces positive gradients for all frames, which is not discriminate enough for positive bags with negative instances. \emph{Linear softmax pooling}~\cite{linear pl}, \emph{exponential softmax pooling}~\cite{exp pl}, and \emph{attention pooling}~\cite{att} assign a different weight to each frame, thereby varying the sign and magnitude of the their gradients. Wang \etal~\cite{five pl} compared the above five pooling functions, and demonstrated that linear softmax pooling was the best at localizing sound events because it could produce positive gradients for some frames and negative gradients for others. McFee \etal~\cite{auto pl} developed a family of adaptive pooling operators named \emph{auto-pool} which could achieve a similar effect. He \etal~\cite{hier pl} proposed a hierarchical pooling structure.

It remains challenging, however, to predict the onsets and offsets of sound events using these pooling functions. A core reason is that environmental sounds in general have less structure in comparison to speech and music. Many independent sources (\textit{e.g.} animals, vehicles, electrical appliances) entail characteristics with considerable variability. One of the key factors that cause variability is the various durations of different events, and these pooling functions cannot adapt to such variability. Besides, as synthetic strongly labelled data and unlabelled data are easier to obtain in real life, studies and challenges such as DCASE 2019 Task~4~\cite{DCASE2019} have turned to semi-supervised SED with synthetic strongly labelled, weakly labelled and unlabelled data. Nevertheless, to our knowledge, few studies have investigated the effects of applying different pooling functions in semi-supervised SED.

To mitigate the above issues, we design a simple but effective pooling function termed as \emph{power pooling}. We use a power function of the predicted frame-level probability as the weight for each frame, and set the exponent as a trainable parameter to automatically generate an optimal threshold between positive and negative gradients. The power pooling provides variable gradient directions for frame-level predictions with a flexible threshold and improves its adaptivity. The trainable power parameter can also be made dependent on the event category, thus adapting to sources with different types of acoustic characteristics. We evaluate the proposed method on a purely weakly labelled dataset (DCASE 2017 Task~4) and a semi-supervised dataset (DCASE 2019 Task~4). Our empirical results show that power pooling outperforms other pooling functions on all metrics.

\section{Pooling functions}

Polyphonic SED with $C$ event categories can be regarded as $C$ binary MIL problems. As such, we will only consider one event category hereafter. Each training clip can be regarded as a bag $(X,Y)$, where $X = \{x_1, \ldots, x_m\}$ are the instances (frames), and $Y \in \{0,1\}$ is the label of the bag. As illustrated in Fig.~\ref{fig:intro}, an SED system predicts a frame-level probability $y_i^f \in [0,1]$ for each instance $x_i$. The pooling function aggregates $y_i^f$ into the clip-level prediction $y^c \in [0,1]$ by assigning a weight $w_i$ to each frame and taking the weighted average:
\begin{equation}\label{equ:forward}
  y^c = \frac{\sum_{i}y_i^f \times w_i}{\sum_{i}w_i}.
\end{equation}
The loss function is chosen to be the cross-entropy between the clip-level prediction $y^c$ and the label $Y$. During back-propagation, the pooling function determines the gradient received by each instance, and the gradients should have appropriate signs. The formulas of the gradients for five types of pooling functions can be found in~\cite{five pl}.

A pooling function should generally assign larger weights to frames with larger predicted probabilities. This is in order to conform to the \emph{standard multiple instance} (SMI) \emph{assumption}: the bag label is positive if and only if the bag contains at least one positive instance. The state-of-the-art linear softmax pooling uses $y_i^f$ itself as the weight $w_i$ :
\begin{equation}\label{equ:linear}
  y^c = \frac{\sum_{i}(y_i^f \times y_i^f)}{\sum_{i} y_i^f},
\end{equation}
and its gradient is
\begin{equation}\label{equ:linear gradient}
  \frac{\partial y^c}{\partial y_i^f} = \frac{ 2 \times y_i^f - y^c}{\sum_{j} y_j^f}.
\end{equation}

\begin{table}[h]
\processtable{The directions of the clip-level and frame-level gradients in linear softmax pooling ($\theta = 1/2$) and power pooling ($\theta = n/(n+1)$).}
{\label{linear}}
{\begin{tabular}{|l|l|l|l|l|l|l|}
    \hline
    Label                             & Clip-level                      & Condition                    & Frame-level     \\
    \hline
    \multirow{2}{*}{positive ($Y=1$)} & \multirow{2}{*}{$y^c\rightarrow1$}  &$ y_i^f > \theta \times y^c $ & $y_i^f\rightarrow1$ \\
                                      &                                       &$ y_i^f < \theta \times y^c$  & $y_i^f\rightarrow0$   \\
    \hline
    \multirow{2}{*}{negative ($Y=0$)} & \multirow{2}{*}{$y^c\rightarrow0$}  &$ y_f>\theta \times y^c$      & $y_i^f\rightarrow\theta \times y^c$  \\
                                      &                                     &$ y_f < \theta \times y^c$    & $y_i^f\rightarrow \theta \times y^c$  \\
    \hline
  \end{tabular}}
\end{table}

This gradient is positive if and only if $y_i^f > y^c / 2$. For positive clips ($Y = 1$), this causes ``larger'' frame-level probabilities to increase and ``smaller'' frame-level probabilities to decrease, thereby yielding clear boundaries of event occurrences. The threshold between ``larger'' and ``smaller'' probabilities is given by $d = y^c / 2$. For negative clips ($Y = 0$), the gradients pushes all frame-level probabilities toward $y^c / 2$. Considering that this threshold is smaller than $y^c$, all the frame-level probabilities will converge to 0 as desired after enough iterations. The movements of the frame-level probabilities are listed in Table~\ref{linear} as well as depicted in Fig.~\ref{fig:intro}.

Define $\theta = d / y^c$ as the ratio between the threshold at which the gradient changes sign and the clip-level predicted probability. In linear softmax pooling, $\theta$ is fixed at $1/2$. In reality, however, it may be desirable to have a different $\theta$ for different event categories. For example, we may want to boost the predicted probabilities of more frames when a clip contains a type of event that usually lasts a long time (\textit{e.g.} vacuum cleaner), and boost fewer frames when the type of event in question is usually transient (\textit{e.g.} dog bark). This motivated us to propose \emph{power pooling}.

\section{Power pooling}

Without changing the pattern of how predictions move in Table~\ref{linear}, we hope to make the threshold $d$ variable by adding a small number of trainable parameters on the basis of the linear softmax pooling function. We use a trainable parameter, $n$, as the exponent of the frame-level probabilities, $y_i^f$, and the formula of pooling function can be written as:
\begin{equation}\label{equ:power}
  y^c = \frac{\sum_{i}y_i^f \times (y_i^f)^n}{\sum_{i}(y_i^f)^{n}},
\end{equation}
and its gradient can be written as:
\begin{equation}\label{equ:power gradient}
  \frac{\partial y^c}{\partial y_i^f} = \frac{(n+1) \times(y_i^f)^n -n\times (y_i^f)^{n-1} \times y^c}{\sum_{j}(y_j^f)^n}.
\end{equation}
To conform to the SMI assumption, $w_i = (y_i^f)^n$ must be an increasing function, therefore $n$ must be non-negative. The gradient will still have different signs for different frames; the threshold is given by \mbox{$d = \theta \cdot y^c$}, where $\theta = n/(n+1) \in (0,1)$. During back-propagation, frame-level probabilities will move in the same pattern as in Table~\ref{linear}. The pooling takes the $n$-th power of the frame-level probability $y_i^f$ as its weight, so we refer to it as \emph{power pooling}.

Power pooling inherits the advantage of linear softmax pooling at localizing sound events. In addition, the learnable power parameter allows it to approach either max pooling (as $n \rightarrow +\infty$) or average pooling (as $n \rightarrow 0$). When $n$ is fixed to 1, power pooling reduces to linear softmax pooling. It is desirable to make the power $n$ depend on the event category: for long-lasting events we prefer to choose a smaller $n$, which results in a lower threshold and boosts more frames; for transient events we would do the opposite. When the power $n$ gets too large, however, power pooling can suffer from the same problem of zero gradients as max pooling. To avoid this, we add a regularization term of $\lambda \sum_c n_c^2$ to the loss function, where $n_c$ is the power parameter for event category $c$.

\section{Experiments and discussion}

We carry out experiments to compare the performance of power pooling with other pooling functions on a purely weakly labelled dataset---DCASE 2017 Task~4~\cite{data2017} and a semi-supervised SED dataset---DCASE 2019~\cite{DCASE2019}. The realistic recordings of both datasets are subsets of the AudioSet dataset~\cite{audioset}, and DCASE 2019 has a subset with synthetic recordings. Most clips have a duration of 10~seconds (a few clips are shorter), and multiple audio events may occur at the same time.

The dataset of Task~4 of the DCASE 2017 challenge is composed of 17 types of ``warning'' and ``vehicle'' sounds. We take the weakly labelled training set (51,172 clips) and the strongly labelled public test set (488 clips). The DCASE 2019 dataset focuses on 10 types of sound events in domestic environments. It consists of three training sets (synthetic strongly labelled: 2,045 clips, weakly labelled: 1,578 clips, unlabelled: 14,412 clips) and a validation set (1,168 clips). The mean durations of events in DCASE 2017 are in the range of 4-10~s, covering more than 40\% of the clips. The mean durations in DCASE 2019 are gathered in 0.5-5~s, covering less than 50\% of the clips. These two datasets contain relatively long and short events, respectively.

The performance of systems is measured with fine-grained and coarse-grained (clip-level) $F_1$, which balances precision and recall. For fine-grained evaluation, we adopt both event-level metrics with a 200~ms collar on onsets and a collar of 200~ms or 20$\%$ of the event length on offsets, and segment-level metrics with the segment duration set to 1~s. Besides, we aggregate the metrics across event categories using the macro-average. The evaluation details can be found in~\cite{evaluation}.

The data preprocessing and model architecture on DCASE 2017 are nearly the same as in~\cite{five pl}. In a nutshell, the input filterbank features have 400 frames and 64 frequency bins, and the model structure consists of 3~convolutional blocks, 2~BiGRU layers and 1~dense layer. We add a batch norm layer to each convolutional block. Since DCASE 2019 contains unlabelled data, a semi-supervised framework is applied. The data preprocessing and backbone model are based on methods in~\cite{mt18}. We adopt the popular \emph{mean-teacher}~\cite{MT} architecture and a feature extractor with convolutional recurrent neural networks~\cite{CRNN}. Furthermore, the following optimizations are performed: First, we augment the data by shifting input features along the time axis, sampling the shift from a normal distribution with zero mean and a standard deviation of 16~frames. Second, we adopt the architecture of the feature extractor in~\cite{mt19}. Third, we apply a set of median filters on the frame-level predicted probabilities, using window sizes proportional to the average duration of each event category. Finally, the regularization hyperparameter $\lambda$ for the power parameters is set to $10^{-4}$ for DCASE 2017 and 0 for DCASE 2019.

\begin{table}[h]
\processtable{Detailed results on DCASE 2017. }
{\label{2017}}
{\setlength{\tabcolsep}{3.0pt}\begin{tabular}{|c|c|c|c|c|c|c|c|}
\hline
\multirow{3}{*}{Pooling Function}   &\multicolumn{7}{c|}{DCASE 2017}   \\
\cline{2-8}
					&\multicolumn{3}{c|}{Event-level}  &\multicolumn{3}{c|}{Segment-level}  &Clip-level \\
\cline{2-8}
                   & $F_1$             & Precision   &Recall        & $F_1$          &Precision   &Recall        & $F_1$\\
\hline
Max                & 0.094             &0.169        &0.073        & 0.372           & 0.567      &0.300          & 0.465 \\
Average            & 0.165             &0.147        &0.196        & 0.450           & 0.425      &0.515          & 0.516 \\
Exponential        & 0.166             &0.154        &0.187        & 0.466           & 0.472      &0.481          & 0.521 \\
Attention~\cite{att} & 0.130             &0.139        &0.171        & 0.434           & 0.481      &0.470          & 0.527 \\

\hline
Auto~\cite{auto pl}& 0.169             &0.144        &0.217        & 0.457           & 0.425      &0.535          & 0.536 \\
CAP~\cite{auto pl} & 0.164             &0.152        &0.199        & 0.468           & 0.447      &0.521          & \textbf{0.544} \\
RAP $10^{-2}$~\cite{auto pl} & \textbf{0.176}    &0.147        &0.233        & 0.464           & 0.411      &0.561          & 0.532 \\
RAP $10^{-3}$~\cite{auto pl} &0.165              &0.145        &0.202        & 0.464           & 0.432      &0.529          & 0.534 \\
RAP $10^{-4}$~\cite{auto pl} & 0.158             &0.132        &0.206        & 0.455           & 0.410      &0.539          & 0.526 \\

\hline
Linear~\cite{five pl} & 0.162             &0.178        &0.161        & \textbf{0.471}  & 0.542      &0.451          & 0.535 \\

\hline
Power              & \textbf{0.196}    &0.168        &0.248        & \textbf{0.480}  & 0.460      &0.537          & \textbf{0.545} \\
\hline
\end{tabular}}
\end{table}

\begin{table}[h]
\processtable{Detailed results on DCASE 2019.}
{\label{2019}}
{\setlength{\tabcolsep}{3.0pt}\begin{tabular}{|c|c|c|c|c|c|c|c|}
\hline
\multirow{3}{*}{Pooling Function}  &\multicolumn{7}{c|}{DCASE 2019} \\
\cline{2-8}
					&\multicolumn{3}{c|}{Event-level}  &\multicolumn{3}{c|}{Segment-level}  &Clip-level \\
\cline{2-8}
                   & $F_1$             & Precision   &Recall        & $F_1$          &Precision   &Recall        & $F_1$\\
\hline
Max               & 0.256             &0.381        &0.201        & 0.488           & 0.836      &0.362          & 0.609\\
Average           & 0.171             &0.158        &0.201        & 0.564           & 0.499      &0.675          & 0.597\\
Exponential       & 0.187             &0.190        &0.203        & 0.569           & 0.559      &0.654          & 0.611\\
Attention~\cite{att}         & 0.320             &0.359        &0.300        & \textbf{0.600}  & 0.688      &0.547          & 0.386\\
\hline
Auto~\cite{auto pl}              & 0.218             &0.288        &0.180        & 0.597           & 0.755      &0.526          & \textbf{0.655}\\
CAP~\cite{auto pl}               & 0.188             &0.217        &0.171        & 0.598           & 0.628      &0.601          & 0.641\\
RAP $10^{-2}$~\cite{auto pl}     & 0.177             &0.189        &0.179        & 0.584           & 0.544      &0.668          & 0.639\\
RAP $10^{-3}$~\cite{auto pl}     & 0.172             &0.175        &0.181        & 0.586           & 0.530      &0.682          & 0.640\\
RAP $10^{-4}$~\cite{auto pl}     & 0.178             &0.229        &0.151        & 0.537           & 0.602      &0.533          & 0.516\\
\hline
Linear~\cite{five pl}            & \textbf{0.343}    &0.431        &0.292        & 0.583           & 0.738      &0.498          & \textbf{0.655} \\
\hline
Power             & \textbf{0.378}    &0.437        &0.340      & \textbf{0.624}    &0.752       &0.547          & \textbf{0.694}\\
\hline
\end{tabular}}
\end{table}

Table~\ref{2017} and Table~\ref{2019} compare power pooling with three classic pooling functions, the popular attention pooling~\cite{att}, a family of auto pooling (Auto, CAP, RAP)~\cite{auto pl} and the baseline linear softmax pooling~\cite{five pl} on DCASE 2017 and DCASE 2019 datasets. We indicate in bold the best $F_1$ scores across all systems, as well as the best apart from power pooling. The impact of pooling functions on the semi-supervised dataset and the weakly labelled dataset appears to be similar. Power pooling achieves the highest $F_1$ at all the three levels on both datasets. The fact that power pooling benefits both fine-grained and coarse-grained SED indicates that it yields proper weights and gradients. As for event-level SED which this article focuses on, power pooling shows an improvement of 2\% and 3.5\% absolute, or 11.4\% and 10.2\% relative on the two datasets. The baseline linear softmax pooling function produces significantly worse recall scores than power pooling, while achieving only comparable precision. This phenomenon demonstrates that the power pooling can find better power parameters and reduce false negatives.

\begin{figure}[ht]
\centering{\includegraphics[width=80mm]{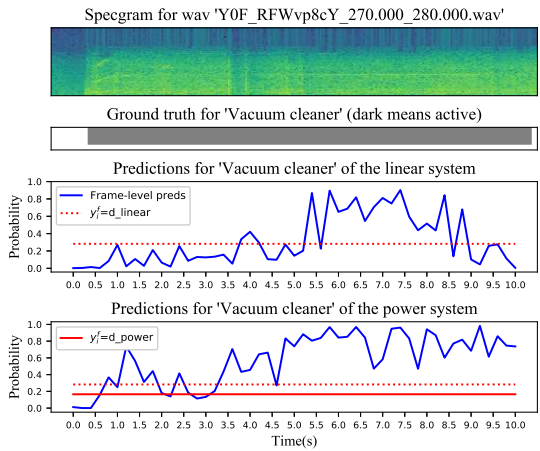}}
\caption{Frame-level predictions of linear softmax and power pooling at epoch~40.}
\label{fig:preds}
\source{}
\end{figure}

Fig.~\ref{fig:preds} illustrates the predictions for the ``vacuum cleaner'' event on a weakly labelled training clip, produced by a system after 40~epochs of training (we trained for 200~epochs in total). We also show the actual temporal interval spanned by the event. For the power pooling system, we show the threshold between positive and negative gradients arising from both the power pooling function ($d_\text{power} = n/(n+1) \cdot y^c$) and the linear softmax pooling function ($d_\text{linear} = 1/2 \cdot y^c$). The power parameter for the ``vacuum cleaner'' event is $n = 0.337 < 1$; therefore power pooling yields a lower threshold, allowing more frames to receive positive gradients. Compared with the linear softmax pooling system, more frames in the power pooling system receive a gradient in the correct direction, notably from 0.5~s to 3~s. This indicates how a more appropriate threshold between positive and negative gradients can help to correctly pinpoint the onset and offset of events.

\begin{figure}[ht]
\centering{\includegraphics[width=80mm]{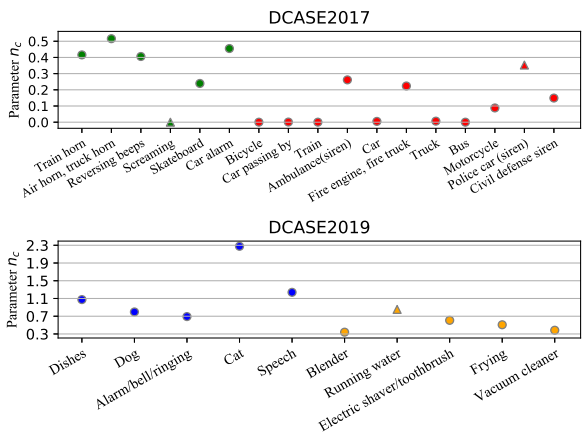}}
\caption{The power parameter $n_c$ for each event category in the final model.}
\label{fig:class_weight}
\source{}
\end{figure}

Fig.~\ref{fig:class_weight} shows the power parameters $n_c$ of each event category in the final model. We sort the event categories horizontally by their average duration, and divide them roughly into shorter events (green and blue symbols) and longer events (red or yellow symbols). The durations of shorter and longer events in the DCASE 2017 dataset fall within 4--5~s and 6--10~s; for DCASE 2019, these durations fall within 0.5--1.1~s and 2--5~s. Except for a small number of categories (represented by triangles), longer events tend to have smaller power parameters, making the power pooling approach average pooling; shorter events tend to have larger power parameters, making the power pooling approach max pooling. This is observed both within each dataset and across the two datasets, and agrees well with the motivation.

\section{Conclusion}

This paper has proposed a practical power pooling function for weakly labelled SED. Power pooling overcomes the shortcoming of the state-of-the-art linear softmax pooling that the weight of a frame is determined by a fixed formula from its predicted probability. With only one learnable power parameter per event category added, power pooling can automatically learn an appropriate threshold between positive and negative gradients for each event category. This allows power pooling to adapt to various sound events of different time scales. Experiments illustrate that power pooling achieves the highest $F_1$ on all levels on both weakly labelled SED and semi-supervised SED. Moreover, the power pooling function is generic enough to be applied to MIL problems in other domains.

\vskip3pt
\ack{This work is supported by the National Natural Science Foundation of China (No. 62071461).}

\vskip5pt

\noindent Yuzhuo Liu, Hanting Chen and Pengyuan Zhang(\textit{Key Laboratory of Speech Acoustics \& Content Understanding, Institute of Acoustics, University of Chinese Academy of Sciences, China})
\vskip3pt
\noindent Yun Wang(\textit{Facebook AI Applied Research})
\vskip3pt

\noindent E-mail: liuyuzhuo@hccl.ioa.ac.cn, chenhangting@hccl.ioa.ac.cn, yunwang@fb.com, zhangpengyuan@hccl.ioa.ac.cn

\end{document}